\documentclass[12pt]{amsart}
\usepackage{epsfig}

\title[propagation of semiclassical wigner functions]{On the propagation of semiclassical Wigner functions}

\author[Rios \& Ozorio]{P.P. de M. Rios \ \& \ A.M. Ozorio de Almeida}

\address{{\it Rios}: Laborat\'orio Nacional de Computa\c c\~ao
Cient\'{\i}fica, 
Av. Getulio Vargas 333, Petr\'opolis, RJ,  25651-070, Brasil.  
{\it Present address}: Department of Mathematics, University of California, Berkeley, CA, 94720-3840, USA.}
\email{prios@math.berkeley.edu}

\address{{\it Ozorio}: Centro Brasileiro de Pesquisas F\'{\i}sicas, Rua Dr. Xavier 
Sigaud 150, Urca, 
Rio de Janeiro, RJ, 22290-180, Brasil. {\it Present address}: Max Planck Institute for Physics of Complex Systems, 
Noethnitzer Str.38, 0118, Dresden, Germany.}
\email{ozorio@cbpf.br \  ozorio@mpipks-dresden.mpg.de} 

 1

\def\R{{I\!\!R}}
\def\d{{\partial}}

\begin{document}

\thispagestyle{empty}

\vspace{1cm}

\begin{abstract}

We establish the difference between the propagation
of semiclassical Wigner functions and classical Liouville propagation.
First we re-discuss the semiclassical limit for the propagator of Wigner functions, 
which on its own leads to their classical propagation. Then, 
via stationary phase evaluation of the full integral evolution equation, 
using the semiclassical expressions of Wigner functions, 
we provide the correct geometrical prescription for their semiclassical propagation. 
This is determined by the classical trajectories of the tips
of the chords defined by the initial semiclassical Wigner function and centered on their 
arguments, in contrast to the Liouville propagation which is determined by the 
classical trajectories of the arguments themselves.

\end{abstract}

\maketitle 
 

\section{introduction}

The Wigner \cite{Wig} function $W$ is a representation of the quantum density operator $\hat{\rho}$ as a real function on phase space, 
for a non-relativistic dynamical system with classical-quantum correspondence and $n$ degrees of 
freedom. Here, all the underlying geometry is euclidean and we can include the Wigner function within the general framework  
of Weyl representation of operators: $\hat{A}\to A\in\mathcal{C}^k_{\R}(\R^{2n})$, so $W(x)=(2\pi\hbar)^{-n}\rho(x)$ , where $x\equiv (p_1,...p_n,q_1,...q_n)$ denotes a point in phase space. 
In this context, the quantum evolution of the density operator $\hat{\rho}_t$ effected by an autonomous external hamiltonian $\hat{H}$ 
can be written in terms of its Weyl \cite{Wey} representation as: 
\begin{equation} \label{ham} 
\d W(x,t)/\d t \ = \{ H , W_t \}(x) + O(\hbar^2) \ , 
\end{equation} 
where $W_t(x)\equiv W(x,t)$ , $H$ is the Weyl symbol of $\hat{H}$ and $\{ \ , \ \}$ is the classical Poisson bracket. 

Even though the Wigner function $W$ is generally nonpositive and thus cannot be taken as a classical probability density (in contrast to 
its lagrangian averages $\int W(p,q) dp$ , etc...) , equation (\ref{ham}) is the source of a somewhat widespread belief that the propagation of $W_t(x)$ 
coincides in the semiclassical limit ($\hbar \to 0^{+}$) with the classical ($\hbar = 0$) propagation of a Liouville probability density. In fact, for a quadratic hamiltonian  
$\hat{H}^{(2)}$ the semiclassical (exact) propagation of $W_t(x)$ is indeed classical. That is, if $x_0 \to x_t$ is the classical hamiltonian 
flow generated by $H^{(2)}$ , then 
\begin{equation} \label{folk} 
W_0(x_0) \longrightarrow W_t(x_t) = W_0(x_0) \ .
\end{equation}
For a general hamiltonian $\hat{H}$ (whose Weyl symbol $H$ does not necessarily coincide with the classical hamiltonian $h$) the classical 
propagation (\ref{folk}) induced by (\ref{ham}) remains a good approximation (semiclassically correct) if $W$ is a fairly smooth function 
(as with the Weyl representation of an observable), in which case the corrections to (\ref{ham}) can be semiclassically ignored.  
Such a smoothness condition can be realized for highly mixed statistical 
states, in which case $W$ will look very much like a classical probability density in phase space and not surprisingly its 
propagation will be nearly classical. 

On the other hand, pure states are not in general represented by smooth Wigner functions.  Indeed, the semiclassical limit $\mathcal{W}$ for the 
Wigner function of a pure state $\Psi$ in one degree of freedom \cite{Ber} is: 
\begin{equation} \label{wig} 
\mathcal{W}(x) \approx \mathcal{A}_{\psi}(x) \ cos\left( S_{\psi}(x)/\hbar - \pi / 4 \right) \ , 
\end{equation} 
where $S_{\psi}(x)$ is the symplectic area between an arc of a (Bohr-Sommerfeld quantized) curve $\psi$     
and its corresponding chord centered at $x$ , and $\mathcal{A}$ is a smooth amplitude function. Such an expression for $\mathcal{W}$ is highly 
oscillatory, specially at this semiclassical limit, and its successive derivatives knock off all the favorable powers of $\hbar$ in the ``corrections'' 
to (\ref{ham}).  
The important exception is when $x$ lies very close to $\psi$ in which 
case $S_{\psi}(x)$ is very close to zero and constant, and $\mathcal{W}(x)$ is locally smooth.  
However, the regions inside a closed leaf $\psi$ , 
where $\mathcal{W}$ is highly oscillatory, are of utmost importance for they neatly
exhibit the nonpositive and thus nonclassical aspect of $W$ which accounts for quantum interference and coherence phenomena. Thus, (\ref{folk}) provides an inadequate description of the propagation of Wigner functions at the semiclassical level. 

Most of these problems with the propagation of semiclassical Wigner functions were already clearly discussed by Heller \cite{Hel} 
in 1976, even before the semiclassical approximations for $W$ were properly developed  \cite{Ber}\cite{Hel1}. 
The asymptotic expansion (\ref{ham}) in $\hbar$ was then partially resummed to obtain an improved propagation, 
though in a not very general context. From another approach, 
starting with the integral expression of the evolution equation, Marinov \cite{Mar} derived in 1991 a  
path integral representation for the propagator of Wigner functions, obtaining its semiclassical limit, which leads to (\ref{folk}). In this paper we   
develop the geometrical explanation of why the semiclassical limit of the propagation of Wigner functions cannot be derived 
from the semiclassical limit of their propagator.  
 
Thus, in $\S 2$ we present the  integral equation for the Wigner evolution, defining its kernel, 
the Wigner propagator, whose stationary phase evaluation necessarily leads to classical propagation.  
Then, in $\S 3$ we recollect the constructions of semiclassical Wigner functions which will permit, in  $\S 4$, 
a correct stationary phase evaluation of the full integral equation for their evolution. 
This leads, in $\S 5$, to a simple geometrical prescription for the semiclassical limit of the 
propagation of a Wigner function in terms of the classical flow of the tips of the chords centered on 
the arguments of the initial semiclassical Wigner function. We conclude in $\S 6$ with a discussion  
on the geometrical meaning of the difference between classical Liouville and semiclassical Wigner propagation.

\section{integral evolution and semiclassical propagator}

The starting point of our analysis is the expression for the product of operators in the center (or Weyl) representation.  Thus, if 
$A(x)$ and $B(x)$ are the center representations of $\hat{A}$ and  $\hat{B}$ , then the center representation for $\hat{A}\hat{B}$ is given \cite{Moy}\cite{OdA} 
by the integral Moyal product:  
\begin{equation} \label{prod}
[AB](x) = \int dx' dx'' A(x') B(x'') exp(i\Delta(x,x',x'')/\hbar) \ , 
\end{equation} 
where $\Delta(x,x',x'') \equiv 2(x\wedge x' + x'\wedge x'' + x''\wedge x)$ is the symplectic area of the triangle with these midpoints. Iterating  (\ref{prod}) we obtain the integral equation for the evolution of Wigner functions as the Weyl transform of the quantum evolution equation $\hat{\rho}_t = \hat{U}_{-t}\hat{\rho}_0\hat{U}_t$ : 
\begin{equation}  \label{tp1}
W_t(x) = \int dx'dx''dx''' U_{-t}(x')W_0(x'')U_t(x''')exp(2i(x\wedge x' + x''\wedge x''')/\hbar)\delta(x-x'+x''-x''') \ , 
\end{equation}
where $U_t(x)$ is the Weyl propagator, i.e. the Weyl transform of the unitary evolution operator $\hat{U}_t$. 
The $\delta$-function prescribes the 4 points in $\R^{2n}$ as vertices of a parallelogram. The phase is twice the area of this parallelogram  (times $\hbar^{-1}$). But this is also the area of any element of a continuous family of quadrilaterals circumscribed to the given parallelogram. In other words, we can  identify the symplectic area of the parallelogram with vertices 
at $(x, x', x'', x''')$ as half of the symplectic area of any quadrilateral whose sides are centered on these points. 
This brings the product rule for three operators in line with that for two operators and, indeed,
the product of any number of operators will depend on the corresponding circumscribed polygon \cite{OdA}. 
Thus, denoting by $\Delta_4 $ the area of a quadrilateral as function of its midpoints, we have that $\Delta_4 (x,x',x'',x''') \equiv 2(x\wedge x' + x''\wedge x''')$ and this function is well defined only on the subset $D^3 \subset (\R^{2n})^4$, isomorphic to $(\R^{2n})^3$, determined by the $\delta$-function. Fixing one (say, the first, $x$) of these points, we obtain a subset $D^2_x \subset (\R^{2n})^3$, isomorphic to $(\R^{2n})^2$. Denoting its induced measure by $d^2_x(x',x'',x''')$ , we can rewrite (\ref{tp1}) as 
\begin{equation} \label{tp2} 
W_t(x) = \int_{D^2_x} d^2_x(x',x'',x''')U_{-t}(x')W_0(x'')U_{t}(x''')exp(i\Delta_4 (x,x',x'',x''')/\hbar)  \ .
\end{equation} 
Similarly, we can reparametrize the parallelogram by identifying: 
$ x'' \equiv x_0$ ,  $x' \equiv (x+x_0)/2 - \mu$ , \linebreak $x''' \equiv (x+x_0)/2 + \mu$ ,  
in which case we can use (\ref{tp1}) to get an expression for the Wigner propagator or kernel $L_t(x_0,x)$ , via its defining formula
$W_t(x) = \int dx_0 L_t(x_0,x) W_0(x_0)$ , in the following form: 
\begin{equation} \label{L} 
L_t(x_0,x) = \int d\mu \ U_{-t}( (x + x_0)/2 - \mu )U_t( ( x + x_0)/2 + \mu )exp(2i(\mu\wedge(x-x_0))/\hbar) \ . 
\end{equation} 

Marinov \cite{Mar} has derived an explicit path integral representation for the Wigner propagator $L_t(x_0,x)$. This can also be achieved by 
introducing in (\ref{L}) the path integral representation \cite{OdA} for the Weyl propagator $U_t(x)$ . For this latter, 
there exists a well-established semiclassical limit \cite{Mar0}\cite{OdA}: 
\begin{equation} \label{scW} 
\mathcal{U}_t(x) \approx \mathcal{B}_{\gamma_t}(x)exp\left\{ i\hbar^{-1}\left[ S_{\gamma_t} - E_{\gamma_t}t\right] (x) \right\} \ , 
\end{equation} 
where $S_{\gamma_t}(x)$ is the symplectic area between the classical trajectory $\gamma_t$ (determined by the Weyl hamiltonian $H$) and the chord 
centered on $x$, with $E_{\gamma_t} = H(\gamma_t)$ being the energy of this trajectory, and where, again, $\mathcal{B}_{\gamma_t}$ is a slow-varying real 
amplitude function. For sufficiently small times we can guarantee the existence of a single trajectory. Eventually there may be bifurcations, in 
which case (\ref{scW}) must be replaced by a sum of similar terms with appropriate Morse indices \cite{OdA}. 

The naive semiclassical limit for the propagation of Wigner functions can be seen as a consequence of trying to get a semiclassical limit for the Wigner propagator (\ref{L}) 
itself by stationary phase, either via its path integral representation \cite{Mar} or directly by using (\ref{scW}). In any case we are 
lead to a semiclassical Wigner propagator which is indeed classical, i.e. of the singular form: 
\begin{equation} \label{naive}
\mathcal{L}_t(x_0,x)  \approx \delta(x_0 - (x)_{-t}) \ ,  
\end{equation} 
where $x \to (x)_{-t}$ is the inverse of the classical hamiltonian flow of $H$. Obviously, (\ref{naive}) implies (\ref{folk}). 

Therefore, in order to obtain the correct semiclassical limit for the propagation of Wigner functions, one must apply stationary phase arguments 
to the full integral equation (\ref{tp1}-\ref{tp2}) for the Wigner evolution, using the correct semiclassical expressions for the Wigner functions 
themselves.

\section{semiclassical wigner functions}

For one degree of freedom, the simple form of the semiclassical Wigner function (\ref{wig}) depends on there existing only a single chord centered 
on each point $x$ , besides the fulfilling of the  
semiclassical condition itself, i.e. the area $S_{\psi}(x)$ being large in comparison to Planck's constant $\hbar$. These conditions  
hold for points not too close to a convex leaf (curve) $\psi$ , outside the cusped triangular curve well in the interior of $\psi$ known as the 
Wigner caustic \cite{Ber}. Inside the caustic there are three chords centered on each point and the semiclassical Wigner function becomes a superposition 
of contributions of the same form as (\ref{wig}), one for each chord and its associated area. Along the Wigner caustic itself, two chords coalesce and 
the amplitude in (\ref{wig}), defined as: 
\begin{equation} \label{amp} 
\mathcal{A}_{\psi}(x) = (2\omega /\pi)\left( 2\pi\hbar | \dot{x}_{+}\wedge\dot{x}_{-} | \right) ^{-1/2} \ , 
\end{equation} 
blows up. Here, $\omega$ is the classical frequency of motion along the curve $\psi$ , whereas $\dot{x}_{\pm}$ are the phase space velocities at the 
tips $x_{\pm}$ of the corresponding chord centered on $x$. The Wigner caustic is hence the locus of the centers of the chords between points on $\psi$ 
with parallel or anti-parallel tangents. The caustic condition thus leads to the vanishing of the skew-product $\dot{x}_{+}\wedge\dot{x}_{-}$ and since 
this also happens when $x$ converges onto $\psi$, 
this latter can be considered as a separate branch of the Wigner caustic, i.e. the breakdown of (\ref{wig}) also takes place on $\psi$ itself. 
Berry \cite{Ber} derived a uniform approximation based on the Airy function that is also oscillatory and asymptotically equivalent to (\ref{wig}) 
inside $\psi$ , rises to a smooth maximum near to this curve and decays exponentially outside. 

The same picture holds for integrable systems 
with $L$ degrees of freedom \cite{OH}. The novelty is that the caustic still has codimension $1$, so the invariant $L$-torus arises as 
a higher singularity or catastrophe of the Wigner caustic. Another important feature is that multiple chords may be centered on points that 
lie arbitrarily close to the invariant torus.  Still, sufficiently far within the energy-shell we retrieve the asymptotic form 
\cite{OH}\cite{OdA0} 
\begin{equation} \label{mwig} 
\mathcal{W}(x) \approx \sum_j A_j(x) cos \{ S_j(x)/\hbar -n_j\pi /4 \} \ , 
\end{equation} 
where again the actions $S_j(x)$ are bounded by the $j$-th chord centered at $x$ and any arc on the quantized torus (again a lagrangian leaf) between 
the chord tips. The amplitudes are now given by 
\begin{equation}  \label{mamp} 
\mathcal{A}_j(x) = (2/\pi)\left[ (2\pi\hbar)^L  | det \{ I^{+}_j , I^{-}_j \}  | \right]^{-1/2}  \ , 
\end{equation} 
and $n_j$ is the signature of the matrix in (\ref{mamp}). Here, the $2L$ action functions $I^{\pm}_j$ are defined in terms of the $L$ action functions $I$ 
as $I^{\pm}_j(x) \equiv I(x \pm \xi_j/2)$. Note that if $L=1$ (\ref{mamp}) is identical with (\ref{amp}). 

No immediate generalization of the semiclassical expression for pure-state wigner function is available for chaotic or general nonintegrable systems. 
However, the superposition of pure Wigner functions in a classically narrow energy range $\epsilon$ determines the spectral 
Wigner function $W(x,E,\epsilon)$ which has the now familiar semiclassical limit \cite{Ber1}\cite{OdA} 
\begin{equation} \label{spcw} 
\mathcal{W}(x,E,\epsilon) \approx \sum_j \mathcal{A}_j(x) e^{-\epsilon t_j/\hbar} cos \left[ S_j(x)/\hbar - \gamma_j \right]  \ , 
\end{equation} 
where again $S_j(x)$ is the symplectic area between a chord centered on $x$ and a classical path between the chord tips. In this case, this is an arc of a classical trajectory along the energy shell , traversed in the time $t_j$. For $x$ close to the energy shell, the chords are all small and there is one arc 
traversed in a small time and a succession of longer arcs winding around the energy shell. Moving the center $x$ leads to the crossing of many 
Wigner caustics where pairs of chords coalesce and disappear or vice versa. Along these caustics, one can again implement uniform approximations which coincide, off-caustics, with the sum (\ref{spcw}) in which  
the amplitude for each chord is  
\begin{equation} \label{spamp}
\mathcal{A}_j(x) = 2^{L+1}\left\{ (2\pi\hbar) (dE/dt_j) | det[1+M_j] | \right\}^{-1/2}  \ ,  
\end{equation} 
where $M_j$ is the matrix for the linearized classical map near the $j$-th trajectory arc on a special $(2L-2)$-dimensional section that is centro-symmetric 
with respect to $x$. 

Thus we have a persistent overall picture: superposition of rapidly oscillating functions with wave vectors $\xi_j = -J[\d S_j/\d x]$ , 
where $J$ is the standard symplectic matrix in $\R^{2L}$, and amplitudes depending only on the relation between the velocity vectors at the tips of each 
respective chord.

\section{wigner evolution: full stationary phase}

Having recollected the general form of the semiclassical expressions for the various kinds of Wigner functions, we are now in a position to correctly approximate  
(\ref{tp1}-\ref{tp2}) by stationary phase. The crucial point here is how to {\it geometrically} interpret formulas (\ref{tp1}-\ref{tp2}) 
in the semiclassical approximation. As  mentioned earlier, the phase of the exponential in the integrand, which is twice the area of the parallelogram 
(times $\hbar^{-1}$), corresponds to the area 
of {\it any} quadrilateral circumscribed to this parallelogram. Furthermore, recall from the previous section that the argument $x$ of the semiclassical Wigner function (\ref{wig},\ref{mwig},\ref{spcw}) corresponds to the center of the chord whose tips lie on the energy shell or the lagrangian leaf $\psi$ corresponding to the quantum state $\Psi$ , so that the phase of its oscillatory factor  corresponds to the symplectic area between $\psi$ and the chord centered on $x$ (minus $\pi /4$). On the other hand, the semiclassical Weyl propagator (\ref{scW}) also has a phase corresponding to 
the symplectic area between the classical trajectory $\gamma_t$ and the chord centered on $x$ (minus $E_{\gamma}t$). Therefore, all the semiclassical terms of the integrand in (\ref{tp1}-\ref{tp2}) have oscillatory factors whose phases depend on the chords centered on their arguments. This suggests interpreting the phase $\Delta_4 $ of the integrator in (\ref{tp1}-\ref{tp2}) as the area of the circumscribed quadrilateral that {\it fits} 
all the pertinent chords discussed above. To be totally consistent, such fitting must take into account all the correct orientations, so that one should first dismember the Wigner function as $\mathcal{W} = (\mathcal{W}^{+} + \mathcal{W}^{-})/2$ , 
where $\mathcal{W}^{\pm}(x) \approx \mathcal{A}(x)exp\{\pm i(S(x)/\hbar - \pi /4)\}$ , and propagate each member separately, then add them up for the total evolution of $\mathcal{W}$. The situation for a single degree of freedom is illustrated in Figure 1.  

\begin{center} 
\vspace{0.7cm}

\epsfig{file=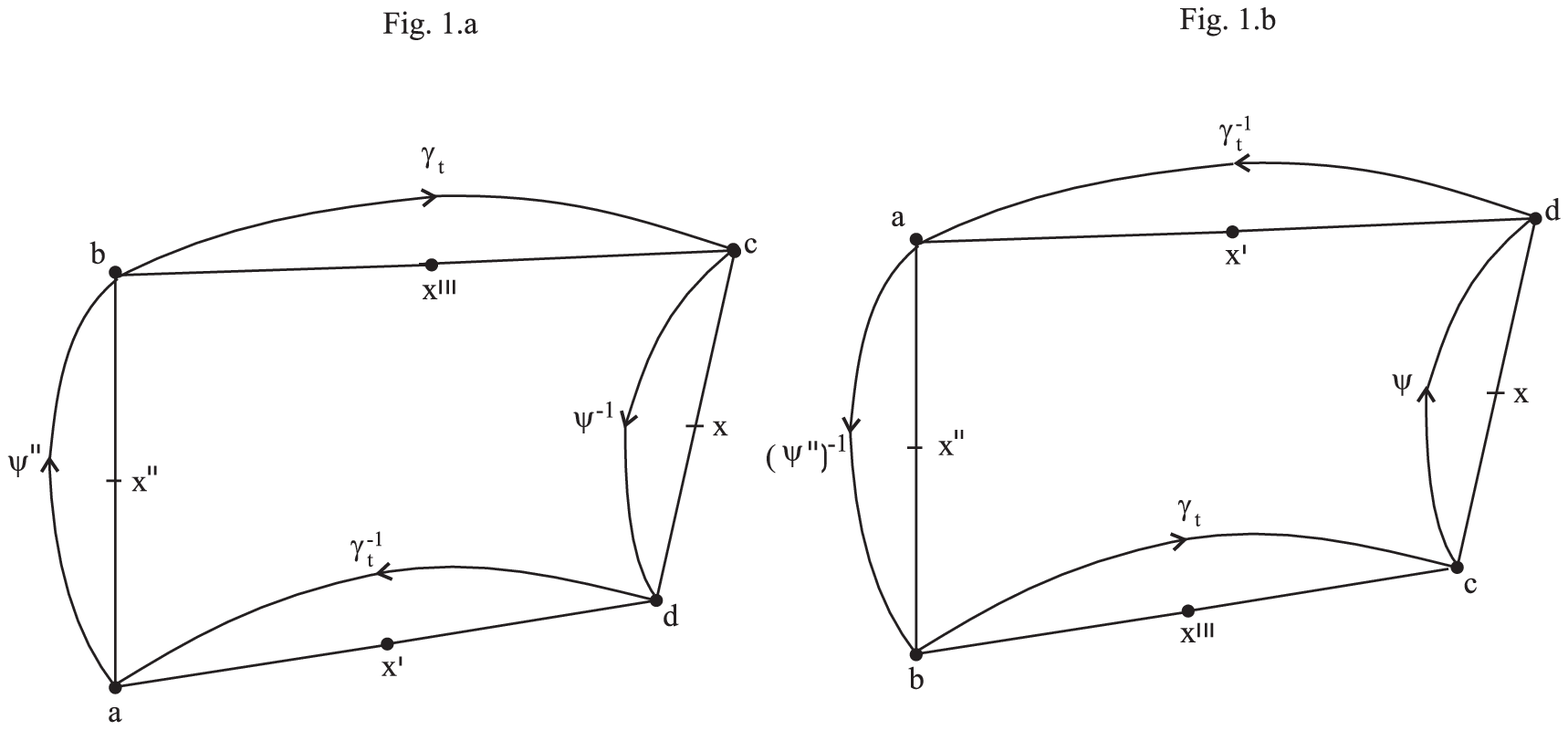,height=7cm} 

\vspace{0.5cm} 

{\footnotesize  Stationary phase condition for the propagation of Wigner functions.}

{\footnotesize Fig. 1.a : Single degree of freedom. The various chords of $\psi''$, $\psi^{-1}$, $\gamma_t$ and $\gamma_{-t}$} 

{\footnotesize centered on $x''$, $x$, $x'''$ and $x'$, respectively, match perfectly with the quadrilateral} 

{\footnotesize  whose sides are centered on these points to yield, via eq.$(16)$, the area between $\psi$} 

{\footnotesize  and the chord centered on $x$, the phase of the propagated Wigner function at $x$.} 

{\footnotesize Fig. 1.b : The same situation but with all the orientations reversed.}  

\vspace{0.7cm}

\end{center} 

The really important issue is that these perfect matchings correspond precisely to the stationary phase condition, as we now show. Let us concentrate on the configuration in Fig.1.a, corresponding to the semiclassical evolution of $\mathcal{W}^{+}_0$, determined by the leaf $\psi ''$, via formulas (\ref{tp1}-\ref{tp2}). The same analysis holds for the other case (Fig.1.b) as well as more dimensions. Inserting expressions (\ref{wig}) and (\ref{scW}) for $\mathcal{W}_0$ and $\mathcal{U}_t$ in (\ref{tp2}), we obtain: 
\begin{equation} \label{W+}  
W^{+}_t(x) = \int_{D^2_x} d^2_x(x',x'',x''') \mathcal{B}'(x')\mathcal{A}''(x'')\mathcal{B}'''(x''')exp\left\{\frac{i}{\hbar}\Phi^{+}(x,x',x'',x''')\right\} \ , 
\end{equation}  
$$ \Phi^{+}(x,x',x'',x''') = [S_{\gamma_{-t}} + E_{\gamma_{-t}}t](x') + S_{\psi ''}(x'') -\pi\hbar /4  + [S_{\gamma_t} - E_{\gamma_t}t](x''') + \Delta_4 (x,x',x'',x''') \ .  
\vspace{0.2cm} 
$$
The stationary phase condition: 
$ \partial\Phi^{+}/\partial x' = \partial\Phi^{+}/\partial x'' = \partial\Phi^{+}/\partial x''' = 0 \ $ 
implies that 
$$\frac{\partial [S_{\gamma_{-t}} + E_{\gamma_{-t}}t]}{\partial x'} = - \frac{\partial\Delta_4 }{\partial x'}  \ \ , \ \  \frac{\partial S_{\psi ''}}{\partial x''} = - \frac{\partial\Delta_4 }{\partial x''} \ \ , \ \  \frac{\partial [S_{\gamma_t} - E_{\gamma_t}t]}{\partial x'''}  =       -  \frac{\partial\Delta_4 }{\partial x'''} \ \ .$$ 
These equations mean \cite{OdA} that the side of the quadrilateral $\Delta_4 $ whose midpoint is $x''$ coincides with the chord of $S_{\psi ''}$ centered on $x''$, with the same orientation as $\psi ''$ . This is the chord  from $a$ to $b$ in Fig(1.a). Similarly for the other cases, the side of $\Delta_4 $ centered on $x'$ is the chord connecting $d$ to $a$ and the side of $\Delta_4 $ centered on $x'''$ is the chord connecting $b$ to $c$. Therefore, the stationary phase condition coincides with the perfectly matching scenario. Furthermore, note that $(E_{\gamma_t} - E_{\gamma_{-t}})t$ is 
the area of the ``curvilinear'' quadrilateral which is formed by the paths $\psi ''$ , $\gamma_t$ , $\psi^{-1}$ , $\gamma_{-t}$ , where $\psi$ is the immage of $\psi''$ under the hamiltonian flow for a time $t$. It follows by direct geometrical inspection on the perfectly matching configuration that 
\begin{equation} \label{proph}
\Big\{ \ S_{\gamma_{-t}}(x') + S_{\psi ''}(x'') + S_{\gamma_t}(x''') + \Delta_4 (x,x',x'',x''')  - [ E_{\gamma_t} - E_{\gamma_{-t}} ] t \  = \  S_{\psi}(x) \ \Big\}_{match}
\end{equation} 
and thus  $\Phi^{+}_{stat}(x) = S_{\psi}(x) - \pi\hbar /4$ . And of course, $\Phi^{-}_{stat} = -\Phi^{+}_{stat}$ . 

Therefore, we retrieve a new Wigner function $W_t$ whose phase is of the same general form given in (\ref{wig},\ref{mwig},\ref{spcw}) for $\mathcal{W}_t$ , in terms of the new action 
$S_{\psi}(x)$. Of course, this is just as we should expect within the general rules of semiclassical self-consistency in which all integrations are carried out within the stationary phase approximation. It should now be feasible to proceed with the full stationary phase evaluation of (\ref{W+}) for 
each of the different types of Wigner functions (\ref{wig},\ref{mwig},\ref{spcw}) previously described, using the semiclassical expression \cite{Mar0}\cite{OdA} for 
the amplitude $\mathcal{B}$ of the Weyl propagator. Such a thorough verification of semiclassical self-consistency has already been carried out \cite{OdA0} 
for the pure state condition $\hat{\rho}^2 = \hat{\rho}$. It relies on such complicated 
geometrical constructions that we consider more appropriate, at this point, to just assume semiclassical self-consistency for the propagated amplitudes, as well. 
Accordingly, each semiclassically propagated Wigner function $\mathcal{W}_t$ retains its respective form (\ref{wig},\ref{mwig},\ref{spcw}), with its corresponding amplitude function  (\ref{amp},\ref{mamp},\ref{spamp}). 

\section{semiclassical propagation: geometrical prescription}

The beautiful consequence of this natural result is that it is sufficient to propagate the two tips of the chord 
of the original Wigner function $\mathcal{W}_0$ , centered on $x''\equiv x_0 \equiv (x_0^{-} + x_0^{+})/2$ , 
in order to obtain the value of the new Wigner function $\mathcal{W}_t$ at 
$x\equiv \tilde{x}_t \equiv (x_t^{-} + x_t^{+})/2$ . Then, 
using (\ref{proph}) we obtain the new phase $S_{\psi}(x)\equiv S_{\psi_t}(\tilde{x}_t)$ from $S_{\psi''}(x'')\equiv S_{\psi_0}(x_0)$. 
Similarly for the amplitudes, the actions
$\mathcal{S}_{\gamma_t}(x''')\equiv \{ S_{\gamma_t} - E_{\gamma_t}t\} (x''')$ and $\mathcal{S}_{\gamma_{-t}}(x')\equiv \{ S_{\gamma_{-t}} + E_{\gamma_{-t}}t\} (x')$  determine the matrices $M_{+}(x''')$ and $M_{-}(x')$ for the linearized classical maps from $x_0^{+}\equiv x_{+}''=x_{-}'''$ 
to 
$x_{+}'''=x_{+}\equiv \tilde{x}_t^{+}$ and from $\tilde{x}_t^{-}\equiv x_{-}=x_{-}'$ to $x_{+}'=x_{-}''\equiv x_0^{-}$ , respectively, by \cite{OdA} 
\begin{equation} \label{M}  
M_{\pm}(y_{\pm}) = [1- J(\d^2\mathcal{S}_{\gamma_{\pm t}}(y_{\pm})/\d y_{\pm}^2)][1+ J(\d^2\mathcal{S}_{\gamma_{\pm t}}(y_{\pm})/\d y_{\pm}^2)]^{-1}  \ , 
\end{equation}  
where $y_{+} \equiv x'''$, $y_{-} \equiv x'$. Hence, the corresponding phase-space velocity vectors for the tips of the 
propagated chord are obtained from the original ones by 
\begin{equation} \label{popvel} 
\dot{x}_{+} = M_{+}\dot{x}_{+}''  \ \ ; \ \  \dot{x}_{-} = M_{-}^{-1}\dot{x}_{-}'' \ \ , 
\end{equation} 
which gives the new amplitudes $\mathcal{A}_{\psi}(x)\equiv \mathcal{A}_{\psi_t}(\tilde{x}_t)$ from 
$\mathcal{A}_{\psi''}(x'')\equiv \mathcal{A}_{\psi_0}(x_0)$ via their defining equations (\ref{amp},\ref{mamp},\ref{spamp}). 
When the central action $\mathcal{S}_{\gamma_t}(y_{+})$ or $\mathcal{S}_{\gamma_{-t}}(y_{-})$ goes through a caustic, one may rely on the chord actions 
$\tilde{\mathcal{S}}_{\gamma_t}(\xi_{+})$ or $\tilde{\mathcal{S}}_{\gamma_{-t}}(\xi_{-})$ instead, for $\xi_{\pm} = \pm (x_t^{\pm} - x_0^{\pm})$, 
where the chord actions are the Legendre 
transform of the central actions: $\tilde{\mathcal{S}}(\xi)\equiv \mathcal{F}(y(\xi),\xi) , \mathcal{F}(y,\xi)=\xi\wedge y - \mathcal{S}(y)$ , to get \cite{OdA}  
\begin{equation} \label{M2} 
M_{\pm}(\xi_{\pm}) = - [1 + J(\d^2\tilde{\mathcal{S}}_{\gamma_{\pm t}}(\xi_{\pm})/\d \xi_{\pm}^2)][1 - J(\d^2\tilde{\mathcal{S}}_{\gamma_{\pm t}}(\xi_{\pm})/\d \xi_{\pm}^2)]^{-1}  \ .  
\end{equation} 
Note, however, that in this case one must be very careful in correctly counting the Maslov indices.  

Therefore, our analysis clearly leads to interpreting the correct point-to-point propagation of a semiclassical Wigner function  
as given by the simple geometrical picture: 
\vspace{0.1cm} 
\begin{equation} \label{corr} 
\mathcal{W}_0(x_0) \longrightarrow \mathcal{W}_t(\tilde{x}_t)   
\vspace{0.1cm}
\end{equation} 
where $\tilde{x}_t$ is the midpoint of $(x_t^{-},x_t^{+})$ , while $(x_0^{-},x_0^{+})$ stands for the 
tips of the original chord centered on $x_0$ . This prescription based on the tips-of-the-chord flow 
provides a precise semiclassical evaluation of $\mathcal{W}_t(\tilde{x}_t)$ from $\mathcal{W}_0(x_0)$.  
Specifically, (\ref{proph}) is used to propagate the phase of $\mathcal{W}$ along the path $x_0 \to \tilde{x}_t$ . For  short 
times, the smooth amplitude may be taken as approximately constant along this path, 
as illustrated below. For longer times the amplitude must also be properly propagated along this path via (\ref{M}-\ref{M2}). In this way, (\ref{corr}) is straightforwardly used to obtain a quantitatively precise semiclassical propagation of $\mathcal{W}$ ``mostly everywhere''. 
But, wherever $\mathcal{W}_t$ 
goes through an inner Wigner caustic, its  
analysis  must be carefully completed by uniform approximations. Even then (\ref{corr}) provides 
most valuable information because (\ref{M}-\ref{M2}) can be used to determine these inner Wigner caustic points 
and (\ref{proph}) determine the contributing new phases which, 
from the knowledge of the corresponding caustics and uniform approximations for $\mathcal{W}_0$  and 
after a careful qualitative analysis of the flow $\psi_0 \to \psi_t$ , 
can lead to the correct (Airy, Pearcey...\cite{Ber}) functions representing $\mathcal{W}_t$ in 
these regions. The on-shell Wigner caustics are quite simpler to deal with 
because there $x_t^{-}\equiv x_t^{+}$ and (\ref{corr}) coincides with (\ref{folk}).   
Thus, we can say that (\ref{corr}) is quantitatively precise ``mostly 
everywhere'' and qualitatively precise everywhere, at the semiclassical level.

As a simple partial illustration of the method, consider the hamiltonian $h = (q^2 + p^2)^2/4$. Operator ordering is not relevant, modulo a constant, 
and $H = h = r^4/4$ .  The trajectories are circles around the origin, but  $\dot{\theta} = r^2$ so the classical flow of $x_0 = (r_0 , \theta_0 \equiv 0)$ is  $x_t = (r_t=r_0 , \theta_t= r_0^2t)$.  
If $x_0$ is the center of the chord whose tips, lying on $\psi_0$ , are $x_0^{-} = (r_{-},\alpha)$ , $x_0^{+} = (r_{+},-\beta)$, then the new center $\tilde{x}_t = (\tilde{r}_t,\tilde{\theta}_t)$ of $(x_t^{-} , x_t^{+})$ is given by 
$ \ \tilde{r}_t^2 = (r_{-}^2 + r_{+}^2)/4 + r_{-}r_{+}cos((r_{+}^2 - r_{-}^2)t - (\alpha + \beta))/2  \ $  ,  $ \tilde{\theta}_t = r_{-}^2t + \alpha + \alpha '\ $ ,
with 
$ 2\tilde{r}_tcos(\alpha ') = r_{-} + r_{+}cos((r_{+}^2 - r_{-}^2)t - (\alpha + \beta))$ . Thus, not only $\tilde{\theta}_t \neq \theta_t$ , but also $ \tilde{r}_t^2 - r_t^2 \equiv  \tilde{r}_t^2 - r_0^2  = r_{+}r_{-}\{cos ((r_{+}^2 - r_{-}^2)t - (\alpha + \beta)) - cos(\alpha + \beta)\}/2$. 

Despite the oddity of this flow, $\mathcal{W}_t(\tilde{x}_t)$ is mostly everywhere obtained from $\mathcal{W}_0(x_0)$ in a straightforward manner from the knowledge of $x_t^{\pm}$ . 
Hence, modulo nontrivial Maslov changes, the phase difference $\delta S(x_0,\tilde{x}_t) = S_{\psi_t}(\tilde{x}_t) - S_{\psi_0}(x_0)$ independs on 
$\psi_0$ or $\mathcal{W}_0$ and is (via (\ref{proph})) given by:
\begin{equation}  \label{difase} 
\delta S(x_0,\tilde{x}_t) = t(r_{+}^4 - r_{-}^4)/4 + r_{+}r_{-}\{ sin(t(r_{+}^2 - r_{-}^2)/2)cos(t(r_{+}^2 - r_{-}^2)/2 - (\alpha + \beta))\}  \ . 
\end{equation} 
On the other hand, the new amplitude depends on the velocities of the tips of the chord determined by $\psi_0$ and thus depends on $\psi_0$ and $\mathcal{W}_0$ 
itself in an intrinsic way. However, for short enough times such that $\tilde{x}_t$ is not too different from $x_t$ , we can approximate the slow-varying 
amplitude by $\mathcal{A}_t(\tilde{x}_t) \approx \mathcal{A}_t(x_t) \approx \mathcal{A}_0(x_0)$ , ``using'' (\ref{folk}) in a more justified way. 
Together with  (\ref{difase})  this gives a first approximation for the semiclassical propagation of {\it any} Wigner function, mostly everywhere and for short times, 
under the $r^4$ hamiltonian. For longer times, 
we obtain  $\mathcal{A}_t(\tilde{x}_t)$ from  $\mathcal{A}_0(x_0)$ via (\ref{amp}) and  (\ref{popvel}), with $M_{\pm}$ determined by (\ref{M}) from 
$\mathcal{S}_{\pm t} = \pm r_{\pm}^2[r_{\pm}^2t -2sin(r_{\pm}^2t)]/4$ with $|y_{\pm}|^2 = r_{\pm}^2cos^2(r_{\pm}^2t/2)$, 
and by (\ref{M2}) from the corresponding $\tilde{\mathcal{S}}_{\pm t}$ .  
As for formula (\ref{folk}), again the precise error will depend on $\psi_0$  and $x_0$ , but if $x_0$ is far from the leaf $\psi_0$ , 
we can approximate $S_{\psi_t}(x_t) - S_{\psi_0}(x_0) \propto t(r_{+} - r_{-})^3$. Thus, for 
any pair $(\psi_0,x_0)$ such that $r_{+}$ and $r_{-}$ differ significantly, (\ref{folk}) can be considerably wrong even for a short time propagation.

\section{conclusion}

We have shown that the propagation of a semiclassical Wigner function 
is correctly determined by the classical flow of the {\it tips} of the chord whose center is the argument of the initial Wigner function, 
not by the classical flow of the argument itself.
This reveals the irrelevance of the semiclassical limit for the
Wigner propagator. Accordingly, in evaluating the integral equation for the evolution of a 
Wigner function, we found that satisfying the stationary phase condition was
tantamount to matching four areas around an appropriate quadrilateral.
However, by reducing the propagation to a single trajectory, traversed in positive and
negative time, the side of the quadrilateral facing the chord of the
Wigner function shrinks to a point and so it could only be matched by a zero length chord. 
Thus we found that the relevant trajectories for a precise semiclassical propagation  
are explicitly determined by the specific semiclassical Wigner function being propagated.  
    
Indeed, one way to correctly evolve a semiclassical Wigner function  $\mathcal{W}_0$
determined by a leaf $\psi_0$ in phase space is to 
classically evolve $\psi_0$ via the hamiltonian flow: $\psi_0 \to \psi_t$ and then evaluate the new semiclassical Wigner function $\mathcal{W}_t$ at each point from the knowledge of $\psi_t$ via (\ref{wig},\ref{amp}-\ref{spamp}), 
as shown in \cite{BB}. Thus,  
to obtain the value of the new semiclassical Wigner function $\mathcal{W}_t$ at $x$ one needs to know, at the very least, the corresponding chord of $\psi_t$ centered on $x$. In fact, our analysis shows that 
the knowledge  of the flow of each pair of points in  $\psi_0$ is sufficient to determine the 
evolved semiclassical Wigner function $\mathcal{W}_t$ at all points. Though these two prescriptions for the evolution of $\mathcal{W}$ are equivalent, in practice the semiclassical point-to-point  propagation of $\mathcal{W}$ is often simpler to determine, both qualitatively and quantitatively, by (\ref{proph}-\ref{M2}) via tips-of-the-chord flow (\ref{corr}).

On the other hand, a mere knowledge of the hamiltonian flow of an argument of $\mathcal{W}_0$ is not enough to reconstruct the new chord and hence the new value of $\mathcal{W}_t$. The exceptions are either linear flow or when the two tips of the corresponding chord coalesce into its center, in other words, when the argument of $\mathcal{W}_0$ lies on 
the leaf $\psi_0$ . For these points the total chord flow coincides with the center flow, thus (\ref{folk}) is verified in the regions of high amplitude near 
the leaf $\psi_0$, but not otherwise for nonlinear flows. It is easy to see why: the phase of $\mathcal{W}_0(x_0)$ 
corresponds to the symplectic area between $\psi_0$ and the chord centered on $x_0$. Under a nonlinear hamiltonian flow, the symplectic area of this closed circuit is preserved, however the immage of the chord is no longer a straight segment and thus the symplectic area between $\psi_t$ and a new chord 
centered on $x_t$ will generally differ from the corresponding area evaluated at $t=0$. 
This difference can be large, even for short times, if the tips of the initial chord are sufficiently far apart, i.e. when the argument of the initial Wigner function is sufficiently 
far from the initial classical leaf. 
Therefore, generally (\ref{folk}) is simply wrong, 
semiclassically. In sharp contrast, the prescription based on the tips-of-the-chord flow (\ref{corr}) provides a  precise semiclassical 
evaluation of $\mathcal{W}_t(\tilde{x}_t)$ from $\mathcal{W}_0(x_0)$. Only when $\tilde{x}_t \equiv x_t$ do (\ref{folk}) and (\ref{corr}) coincide.

\end{document}